\documentclass[a4paper,11pt]{article}

\makeatletter
\gdef\@fpheader{   }
\gdef\@journal{jhep}

\RequirePackage{amsmath}
\RequirePackage{amssymb}
\RequirePackage{epsfig}
\RequirePackage{CJK}
\RequirePackage{graphicx}
\RequirePackage[numbers,sort&compress]{natbib}
\RequirePackage{color}
\RequirePackage[%dvipdfmx
,colorlinks=true
,urlcolor=blue
,anchorcolor=blue
,citecolor=blue
,filecolor=blue
,linkcolor=blue
,menucolor=blue
,pagecolor=blue
,linktocpage=true
,pdfproducer=medialab
,pdfa=true
,CJKbookmarks
]{hyperref}

\newif\ifnotoc\notocfalse
\newif\ifemailadd\emailaddfalse
\newif\iftoccontinuous\toccontinuousfalse

\def\@subheader{\@empty}
\def\@keywords{\@empty}
\def\@abstract{\@empty}
\def\@xtum{\@empty}
\def\@dedicated{\@empty}
\def\@arxivnumber{\@empty}
\def\@collaboration{\@empty}
\def\@collaborationImg{\@empty}
\def\@proceeding{\@empty}
\def\@preprint{\@empty}

\newcommand{\subheader}[1]{\gdef\@subheader{#1}}
\newcommand{\keywords}[1]{\if!\@keywords!\gdef\@keywords{#1}\else%
\PackageWarningNoLine{\jname}{Keywords already defined.\MessageBreak Ignoring last definition.}\fi}
\renewcommand{\abstract}[1]{\gdef\@abstract{#1}}
\newcommand{\dedicated}[1]{\gdef\@dedicated{#1}}
\newcommand{\arxivnumber}[1]{\gdef\@arxivnumber{#1}}
\newcommand{\proceeding}[1]{\gdef\@proceeding{#1}}
\newcommand{\xtumfont}[1]{\textsc{#1}}
\newcommand{\correctionref}[3]{\gdef\@xtum{\xtumfont{#1} \href{#2}{#3}}}
\newcommand\jname{JHEP}

\newcommand\preprint[1]{\gdef\@preprint{\hfill #1}}

\newcommand\note[2][]{%
\if!#1!%
\stepcounter{footnote}\footnotetext{#2}%
\else%
{\renewcommand\thefootnote{#1}%
\footnotetext{#2}}%
\fi}

\newtoks\auth@toks
\renewcommand{\author}[2][]{%
  \if!#1!%
    \auth@toks=\expandafter{\the\auth@toks#2\ }%
  \else
    \auth@toks=\expandafter{\the\auth@toks#2$^{#1}$\ }%
  \fi
}

\newtoks\affil@toks\newif\ifaffil\affilfalse
\newcommand{\affiliation}[2][]{%
\affiltrue
  \if!#1!%
    \affil@toks=\expandafter{\the\affil@toks{\item[]#2}}%
  \else
    \affil@toks=\expandafter{\the\affil@toks{\item[$^{#1}$]#2}}%
  \fi
}

\newtoks\email@toks\newcounter{email@counter}%
\newcommand{\emailAdd}[1]{%
\emailaddtrue%
\ifnum\theemail@counter>0\email@toks=\expandafter{\the\email@toks, \@email{#1}}%
\else\email@toks=\expandafter{\the\email@toks\@email{#1}}%
\fi\stepcounter{email@counter}}
\newcommand{\@email}[1]{\href{mailto:#1}{\tt #1}}

\newcommand*\collaboration[1]{\gdef\@collaboration{#1}}
\newcommand*\collaborationImg[2][]{\gdef\@collaborationImg{#2}}

\newcommand\afterLogoSpace{\smallskip}
\newcommand\afterSubheaderSpace{\vskip3pt plus 2pt minus 1pt}
\newcommand\afterProceedingsSpace{\vskip21pt plus0.4fil minus15pt}
\newcommand\afterTitleSpace{\vskip23pt plus0.06fil minus13pt}
\newcommand\afterRuleSpace{\vskip23pt plus0.06fil minus13pt}
\newcommand\afterCollaborationSpace{\vskip3pt plus 2pt minus 1pt}
\newcommand\afterCollaborationImgSpace{\vskip3pt plus 2pt minus 1pt}
\newcommand\afterAuthorSpace{\vskip5pt plus4pt minus4pt}
\newcommand\afterAffiliationSpace{\vskip3pt plus3pt}
\newcommand\afterEmailSpace{\vskip16pt plus9pt minus10pt\filbreak}
\newcommand\afterXtumSpace{\par\bigskip}
\newcommand\afterAbstractSpace{\vskip16pt plus9pt minus13pt}
\newcommand\afterKeywordsSpace{\vskip16pt plus9pt minus13pt}
\newcommand\afterArxivSpace{\vskip3pt plus0.01fil minus10pt}
\newcommand\afterDedicatedSpace{\vskip0pt plus0.01fil}
\newcommand\afterTocSpace{\bigskip\medskip}
\newcommand\afterTocRuleSpace{\bigskip\bigskip}
\newlength{\affiliationsSep}
\newcommand\beforetochook{\pagestyle{myplain}\pagenumbering{roman}}

\DeclareFixedFont\trfont{OT1}{phv}{b}{sc}{11}

\renewcommand\maketitle{
\pagestyle{empty}
\thispagestyle{titlepage}
\setcounter{page}{0}
\noindent{\small\scshape\@fpheader}\@preprint\par
\afterLogoSpace
\if!\@subheader!\else\noindent{\trfont{\@subheader}}\fi
\afterSubheaderSpace
\if!\@proceeding!\else\noindent{\sc\@proceeding}\fi
\afterProceedingsSpace
{\LARGE\flushleft\sffamily\bfseries\@title\par}
\afterTitleSpace
\hrule height 1.5\p@%
\afterRuleSpace
\if!\@collaboration!\else
{\Large\bfseries\sffamily\raggedright\@collaboration}\par
\afterCollaborationSpace
\fi
\if!\@collaborationImg!\else
{\normalsize\bfseries\sffamily\raggedright\@collaborationImg}\par
\afterCollaborationImgSpace
\fi
{\bfseries\raggedright\sffamily\the\auth@toks\par}
\afterAuthorSpace
\ifaffil\begin{list}{}{%
\setlength{\leftmargin}{0.28cm}%
\setlength{\labelsep}{0pt}%
\setlength{\itemsep}{\affiliationsSep}%
\setlength{\topsep}{-\parskip}}
\itshape\small%
\the\affil@toks
\end{list}\fi
\afterAffiliationSpace
\ifemailadd %% if emailadd is true
\noindent\hspace{0.28cm}\begin{minipage}[l]{.9\textwidth}
\begin{flushleft}
\textit{E-mail:} \the\email@toks
\end{flushleft}
\end{minipage}
\else %% if emailaddfalse do nothing
\PackageWarningNoLine{\jname}{E-mails are missing.\MessageBreak Plese use \protect\emailAdd\space macro to provide e-mails.}
\fi
\afterEmailSpace
\if!\@xtum!\else\noindent{\@xtum}\afterXtumSpace\fi
\if!\@abstract!\else\noindent{\renewcommand\baselinestretch{.9}\textsc{Abstract:}}\ \@abstract\afterAbstractSpace\fi
\if!\@keywords!\else\noindent{\textsc{Keywords:}} \@keywords\afterKeywordsSpace\fi
\if!\@arxivnumber!\else\noindent{\textsc{ArXiv ePrint:}} \href{http://arxiv.org/abs/\@arxivnumber}{\@arxivnumber}\afterArxivSpace\fi
\if!\@dedicated!\else\vbox{\small\it\raggedleft\@dedicated}\afterDedicatedSpace\fi
\ifnotoc\else
\iftoccontinuous\else\newpage\fi
\beforetochook\hrule
\tableofcontents
\afterTocSpace
\hrule
\afterTocRuleSpace
\fi
\setcounter{footnote}{0}
\pagestyle{myplain}\pagenumbering{arabic}
} % close the \renewcommand\maketitle{

\renewcommand{\baselinestretch}{1.1}\normalsize
\divide\@tempdima\baselineskip
\@tempcnta=\@tempdima
\skip\@mpfootins = \skip\footins
\renewcommand{\@dotsep}{10000}

\newcommand\ps@myplain{
\pagenumbering{arabic}
\renewcommand\@oddfoot{\hfill-- \thepage\ --\hfill}
\renewcommand\@oddhead{}}
\let\ps@plain=\ps@myplain

\newcommand\ps@titlepage{\renewcommand\@oddfoot{}\renewcommand\@oddhead{}}

\numberwithin{equation}{section}

\renewcommand\section{\@startsection{section}{1}{\z@}%
                                   {-3.5ex \@plus -1.3ex \@minus -.7ex}%
                                   {2.3ex \@plus.4ex \@minus .4ex}%
                                   {\normalfont\large\bfseries}}
\renewcommand\subsection{\@startsection{subsection}{2}{\z@}%
                                   {-2.3ex\@plus -1ex \@minus -.5ex}%
                                   {1.2ex \@plus .3ex \@minus .3ex}%
                                   {\normalfont\normalsize\bfseries}}
\renewcommand\subsubsection{\@startsection{subsubsection}{3}{\z@}%
                                   {-2.3ex\@plus -1ex \@minus -.5ex}%
                                   {1ex \@plus .2ex \@minus .2ex}%
                                   {\normalfont\normalsize\bfseries}}
\renewcommand\paragraph{\@startsection{paragraph}{4}{\z@}%
                                   {1.75ex \@plus1ex \@minus.2ex}%
                                   {-1em}%
                                   {\normalfont\normalsize\bfseries}}
\renewcommand\subparagraph{\@startsection{subparagraph}{5}{\parindent}%
                                   {1.75ex \@plus1ex \@minus .2ex}%
                                   {-1em}%
                                   {\normalfont\normalsize\bfseries}}

\def\fnum@figure{\textbf{\figurename\nobreakspace\thefigure}}
\def\fnum@table{\textbf{\tablename\nobreakspace\thetable}}

\long\def\@makecaption#1#2{%
  \vskip\abovecaptionskip
  \sbox\@tempboxa{\small #1. #2}%
  \ifdim \wd\@tempboxa >\hsize
    \small #1. #2\par
  \else
    \global \@minipagefalse
    \hb@xt@\hsize{\hfil\box\@tempboxa\hfil}%
  \fi
  \vskip\belowcaptionskip}

\renewenvironment{thebibliography}[1]{%
\begin{oldthebibliography}{#1}%
\small%
\raggedright%
\setlength{\itemsep}{5pt plus 0.2ex minus 0.05ex}%
}%
{%
\end{oldthebibliography}%
}

\makeatother

\usepackage[T1]{fontenc} % if needed
\usepackage{romannum} 
\usepackage{CJK}

\title{{\boldmath Multipole moment and singular source in  Newtonian gravity and in Einstein gravity}}
\author[a,b,1]{Yu-Zhu Chen,\note{chenyuzhu@nankai.edu.cn}}
\author[b]{Yu-Jie Chen, }
\author[b]{Shi-Lin Li, }
\author[b,2]{Wu-Sheng Dai \note{daiwusheng@tju.edu.cn}}

\affiliation[a]{Theoretical Physics Division, Chern Institute of Mathematics and LPMC, Nankai University, Tianjin 300071, P.R. China}
\affiliation[b]{Department of Physics, Tianjin University, Tianjin 300350, P.R. China}

\abstract{The multipole moments are defined as the multipole expansion coefficients of
the gravitational field at infinity. In Newtonian gravity, the multipole
moments are determined by the source distribution --- the multipole integrals
of the source. In this paper, we show that the multipole moments in general
relativity cannot be determined by the multipole integrals of the source. We
provide the multipole integrals in static axial spacetimes, such as, the
Curzon spacetime. The Curzon spacetime possesses the same multipole integrals
of the source with the Schwarzschild spacetime, while they possess different
multipole moments.
}


\begin{document} %正文开始
\begin{CJK*}{GBK}{song}
\maketitle %生成题目

\flushbottom
%(正文开始) ――――――――――――――――――――――――――――――――――――――――――――――――――
\section{Introduction}

In Newtonian gravity, the multipole moments of the gravitational field are
determined by the multipole integrals of the source. However, in general
relativity, the relation between the multipole moments of the gravitational
field and the multipole integrals of the source is unclear.

The multipole moments in general relativity are first defined by Geroch for
static spacetimes \cite{geroch1970multipole,geroch1970multipole2}.\ The
definition is then generalized to stationary spacetimes by Hansen
\cite{hansen1974multipole}. Multipole moments are also defined by Thorne
\cite{thorne1980multipole}, Beig \cite{beig1981multipole}, and others
\cite{fodor1989multipole}. In ref.
\cite{backdahl2005explicit,backdahl2005static,backdahl2007axisymmetric,backdahl2006calculation}%
, the authors provide serial calculations and discussions of multipole moments
in stationary spacetimes. Multipole moments are related with source integrals
in recent works \cite{gurlebeck2014source,hernandez2016source}. In ref.
\cite{hernandez2016source}, the authors provide a pellucid review about the
definition of multipole moments.

The multipole integrals of the source are rarely discussed in general
relativity. One reason is that most exact solutions of the Einstein equation
are vacuum solutions. Nevertheless, singular sources are discussed in
gravitational collapse problems. A singularity is regarded as a singular
source with infinite density while finite mass \cite{oppenheimer1939continued}%
. In a previous work \cite{chen2017singular}, we provide the quantitative
source in the Schwarzschild spacetime and show that the NUT metric which is
generally considered as non-curvature singularity metric possesses a
Dirac-delta source.

In this paper, we show that there are different spacetimes possess the same
multipole integrals of the source so that the multipole moments of the
gravitational field cannot be determined by the multipole integrals of the source.

Spacetimes considered in this paper are vacuum solutions outside
singularities. Nevertheless, they possess singular sources at singularities.
We provide the singular sources in special static vacuum axial solutions of
the Einstein equation, i.e. the Levi-Civita spacetime and the Curzon
spacetime. The singular sources contribute to the multipole integrals of the source.

The multipole integrals contributed by the singular source can be calculated
directly, so does the multipole expansion of the spacetime at infinity. That
is, we obtain the multipole moments of the gravitational field and the
multipole integrals of the source simultaneously in general relativity. The
Curzon spacetime and the Schwarzschild spacetime possess same multipole
integrals, while they possess different multipole expansions at infinity. This
result shows that the multipole moments of the gravitational field in general
relativity cannot determined by the multipole integrals of the source.

In section 2, we introduce the definition of multipole moments in general
relativity and restate the multipole moments problem in general relativity.
That is, whether the multipole moments of the gravitational field is
determined by the multipole integrals of the source in general relativity. In
section 3, we show that there exist singular sources at singularities in
static axial spacetimes and calculate multipole integrals contributed by the
singular sources. Specifically, we calculate the singular source and multipole
integrals in Newtonian gravity, in Levi-Civita spacetime, and in the Curzon
spacetime. These results are used to discuss the multipole moments problem in
general relativity. In section 4, we provide the conclusion and other open
questions concerning singular sources and multipole moments in general relativity.

\section{Multipole moment in general relativity: the problem}

In this section, we introduce the multipole moments problem in general
relativity. Generally, the problem is whether the multipole moments of the
gravitational field in general relativity is determined by the multipole
integrals of the source or not.

\subsection{Multipole moment in Newtonian gravity}

We begin with the definition of the multipole moments in Newtonian gravity. In
Newtonian gravity, the field $\phi$ satisfies the Possion equation
\begin{equation}
\nabla^{2}\phi\left(  \mathbf{x}\right)  =4\pi\rho\left(  \mathbf{x}\right)  ,
\label{Possion}%
\end{equation}
where $\rho\left(  \mathbf{x}\right)  $ is the density of the source and
gravitational constant $G\equiv1$. With the Green function method, $\phi$ can
be expressed as
\begin{equation}
\phi\left(  \mathbf{x}\right)  =-\int_{V}d^{3}\mathbf{x}^{\prime}\frac
{\rho\left(  \mathbf{x}^{\prime}\right)  }{\left\vert \mathbf{x}%
-\mathbf{x}^{\prime}\right\vert }. \label{Potential.Green}%
\end{equation}
With the expansion at $\mathbf{x}\rightarrow\infty$ or $\mathbf{x}^{\prime
}\rightarrow0$
\begin{equation}
\frac{1}{\left\vert \mathbf{x}-\mathbf{x}^{\prime}\right\vert }=\frac{4\pi
}{2l+1}\sum_{l=0}^{\infty}\sum_{m=-l}^{l}\frac{r^{\prime l}}{r^{l+1}}%
Y_{lm}\left(  \theta,\phi\right)  Y_{lm}^{\ast}\left(  \theta^{\prime}%
,\varphi^{\prime}\right)  \text{ \ \ }\left(  r>r^{\prime}\right)  ,
\label{expansion.sphere}%
\end{equation}
where $\mathbf{x}=\left(  r,\theta,\varphi\right)  $ $\mathbf{x}^{\prime
}=\left(  r^{\prime},\theta^{\prime},\varphi^{\prime}\right)  $,
$Y_{lm}\left(  \theta,\phi\right)  $ are the spherical harmonics and $^{\ast}$
represents the complex conjugate, we have
\begin{equation}
\phi\left(  \mathbf{x}\right)  =-\frac{4\pi}{2l+1}\sum_{l=0}^{\infty}%
\sum_{m=-l}^{l}\frac{Q^{\left(  lm\right)  }}{r^{l+1}}Y_{lm}\left(
\theta,\varphi\right)  \label{Potential.expansion}%
\end{equation}
with
\begin{equation}
Q^{\left(  lm\right)  }\equiv\int_{V}d^{3}\mathbf{x}^{\prime}\rho\left(
\mathbf{x}^{\prime}\right)  r^{\prime l}Y_{lm}^{\ast}\left(  \theta^{\prime
},\varphi^{\prime}\right)  . \label{Multipole.standard}%
\end{equation}
$Q^{\left(  lm\right)  }$ are the multipole moments
\cite{jackson1999classical}.

We can also expand $\frac{1}{\left\vert \mathbf{x}-\mathbf{x}^{\prime
}\right\vert }$ in the cartesian coordinate
\begin{equation}
\frac{1}{\left\vert \mathbf{x}-\mathbf{x}^{\prime}\right\vert }=\sum
_{n=0}^{\infty}\sum_{i_{1}=1}^{3}\ldots\sum_{i_{n}=1}^{3}\frac{\left(
2n-1\right)  !!}{n!}\frac{1}{r^{2n+1}}x^{\prime i_{1}i_{2}\ldots i_{n}%
}x^{_{i_{1}}}x^{_{i_{2}}}\ldots x^{_{i_{n}}}, \label{expansion.cartesina}%
\end{equation}
where $\mathbf{x}=\left(  x^{1},x^{2},x^{3}\right)  $, $\mathbf{x}^{\prime
}=\left(  x^{\prime1},x^{\prime2},x^{\prime3}\right)  $, $r=\sqrt{\left(
x^{1}\right)  ^{2}+\left(  x^{2}\right)  ^{2}+\left(  x^{3}\right)  ^{2}}$,
$\left(  2n-1\right)  !!=\left(  2n-1\right)  \left(  2n-3\right)
\ldots3\cdot1$, and $x^{i_{1}i_{2}\ldots i_{n}}$ is the traceless part of
$x^{_{i_{1}}}x^{_{i_{2}}}\ldots x^{_{i_{n}}}$, for example,
\begin{align*}
x^{j}  &  =x^{j},\\
x^{jk}  &  =x^{j}x^{k}-\frac{1}{3}\delta^{jk}\delta_{mn}x^{m}x^{n}=x^{j}%
x^{k}-\frac{1}{3}\delta^{jk}r^{2}.
\end{align*}
With the expansion (\ref{expansion.cartesina}), we have
\begin{equation}
\phi\left(  \mathbf{x}\right)  =-\sum_{n=0}^{\infty}\frac{\left(  2n-1\right)
!!}{n!}\frac{1}{r^{2n+1}}\sum_{i_{1}=1}^{3}\ldots\sum_{i_{n}=1}^{3}\int%
_{V}d^{3}\mathbf{x}^{\prime}\rho\left(  \mathbf{x}^{\prime}\right)  x^{\prime
i_{1}i_{2}\ldots i_{n}}x^{_{i_{1}}}x^{_{i_{2}}}\ldots x^{_{i_{n}}}%
\end{equation}
with the multipole moments
\begin{equation}
Q^{\left(  i_{1}i_{2}\ldots i_{n}\right)  }\equiv\int_{V}d^{3}\mathbf{x}%
^{\prime}\rho\left(  \mathbf{x}^{\prime}\right)  x^{\prime i_{1}i_{2}\ldots
i_{n}}. \label{Multipole.cartesian}%
\end{equation}
$\frac{1}{r^{l}}x^{i_{1}i_{2}\ldots i_{n}}$ and $Y_{lm}\left(  \theta
,\varphi\right)  $ can be expressed by each other since they are two complete
bases \cite{thorne1980multipole}. That is, the multipole moments defined in
eq. (\ref{Multipole.standard}) and eq. (\ref{Multipole.cartesian}) are
equivalent to each other.

The multipole moments play two roles in eqs. (\ref{Potential.expansion}) and
(\ref{Multipole.standard}). On the one hand, the multipole moments are the
expansion coefficients of the field at infinity. On the other hand, the
multipole moments are the multipole integrals over the source. In this paper,
the expansion coefficients of the gravitational field at infinity are called
the multipole moments of the gravitational field or the multipole moments of
the spacetime, and the multipole integrals of the source are called the
multipole moments of the source. In Newtonian gravity, these two physical
quantities coincide. That is, the gravitational field or the multipole moments
of the spacetime are completely determined by the multipole integrals of the
source or the multipole moments of the source.

\subsection{Multipole moment in general relativity}

In general relativity, the multipole integrals of the source and the multipole
moments of the spacetime are two different physical quantities.

We only consider the static and asymptotically flat metric with the following
form
\begin{equation}
ds^{2}=-g_{00}\left(  x^{k}\right)  dt^{2}+g_{ij}\left(  x^{k}\right)
dx^{i}dx^{j} \label{metric.static}%
\end{equation}
with $i$, $j$, $k=1$, $2$, $3$. The Einstein equation of the metric
(\ref{metric.static}) can be expressed as \cite{hernandez2016source}
\begin{align}
D_{i}D^{i}\xi &  =4\pi\xi\left(  -T_{\text{ }0}^{0}+T_{\text{ }i}^{i}\right)
\equiv4\pi\xi\rho_{M},\nonumber\\
R_{ij}-\frac{1}{\xi}D_{i}D_{j}\xi &  =8\pi\left(  T_{ij}-\frac{1}{2}%
g_{ij}T_{\text{ }\alpha}^{\alpha}\right)  ,\nonumber\\
g_{00}  &  \equiv-\xi^{2}<0 \label{Einstein.static}%
\end{align}
in the orthonormal frame, where $D_{i}$ is the covariant derivative with
respect to the space metric $g_{ij}$, $T_{\text{ }\alpha}^{\alpha}=T_{\text{
}0}^{0}+T_{\text{ }1}^{1}+T_{\text{ }2}^{2}+T_{\text{ }3}^{3}$, $T_{\text{ }%
i}^{i}=T_{\text{ }1}^{1}+T_{\text{ }2}^{2}+T_{\text{ }3}^{3}$, and the
subscript "$M$" of $\rho_{M}$ means "matter".

$\xi$ or $\xi-1$ plays the role of a "Newtonian gravitational potential"
\cite{geroch1970multipole2}. The multipole moments of the spacetime (the
multipole moments of the gravitational field) in general relativity are the
expansion coefficients of $\xi$ at infinity \cite{geroch1970multipole2}. We
remind that the multipole moments of the spacetime are only well defined for
asymptotically flat spacetimes. $g_{00}$ and $g_{ij}$ are not independent
variables since the expansion coefficients of $g_{00}$ and $g_{ij}$ at
infinity can be uniquely determined by the multipole moments of the spacetime
in "asymptotically Cartesian and mass centered" coordinates defined by Thorne
\cite{thorne1980multipole}.

In Newtonian gravity, the gravitational potential satisfies a linear equation.
The expansion coefficients of gravitational potential and the multipole
integrals coincides. That is, the multipole moments of the spacetime is
uniquely determined by the multipole integrals of the source. In general
relativity, $\xi$ is related to the source $\rho_{M}$ in nonlinear equations
(\ref{Einstein.static}). The multipole moments problem is whether the
multipole moments of the spacetime is uniquely determined by the multipole
integrals of the source $\rho_{M}$ or $\xi\rho_{M}$.

The multipole integrals are defined similarly as in Newtonian gravity
\begin{equation}
Q_{M}^{\left(  lm\right)  }=\int dV\rho_{M}r^{l}Y_{lm}^{\ast}\left(
\theta,\varphi\right)  \label{Multipole.Scalar}%
\end{equation}
where $r$ is the radial coordinate, $dV\equiv\sqrt{g}dx^{1}dx^{2}dx^{3}$ and
$g=\xi^{2}\det g_{ij}$. Actually, $r$ is very difficult to define.
Nevertheless, the strict definition of $r$ will not influence the conclusion
in this paper so that we will not concentrate on it. For convenience of
comparison, we generalize the definition (\ref{Multipole.Scalar}) in an
orthonormal frame as a tensor form
\begin{equation}
Q_{\mu\nu}^{\left(  lm\right)  }=\int dVT_{\mu\nu}r^{l}Y_{lm}^{\ast}.
\label{Multipole.Tensor}%
\end{equation}
Strictly speaking, the tensor integrals $Q_{\mu\nu}^{\left(  lm\right)  }$ in
eq. (\ref{Multipole.Tensor}) may be not well defined. Nevertheless, they give
all information in $Q_{M}^{\left(  lm\right)  }$ since $Q_{M}^{\left(
lm\right)  }$ $=$ $-Q_{00}^{\left(  lm\right)  }+Q_{11}^{\left(  lm\right)
}+Q_{22}^{\left(  lm\right)  }+Q_{33}^{\left(  lm\right)  }$.

Now the multipole moments problem can be expressed as whether the multipole
moments of the spacetime can be determined by $Q_{M}^{\left(  lm\right)  }$.

By the way, spacetimes considered in this paper are axisymmetric so that
$T_{\mu\nu}=T_{\mu\nu}\left(  r,\theta\right)  $ (or $T_{\mu\nu}=T_{\mu\nu
}\left(  \rho,z\right)  $) and $Q_{\mu\nu}^{\left(  lm\right)  }=0$ if
$m\neq0$.

\section{Singular source and multipole integral in static axial spacetimes}

The multipole moments problem in general relativity is about the relation
between the multipole expansion of the gravitational field at infinity and the
multipole integrals of the source. The multipole expansion of the
gravitational field at infinity is widely discussed, while the multipole
integrals are rarely calculated.

In a previous work \cite{chen2017singular}, we show that there exists a Dirac
delta source in the Schwarzschild spacetime. In this section, we show that
there exist singular sources at singularities of static axial spacetimes.
Singular sources contribute to multipole integrals. We calculate the singular
source and the multipole integrals in Levi-Civita spacetime and compare with
the previous result to verify the validity of our method. We calculate the
singular source and multipole integrals in the Curzon spacetime and point out
that the Curzon spacetime possesses the same multipole integrals with the
Schwarzschild spacetime. The Curzon spacetime and the Schwarzschild spacetime
possess same multipole integrals, which is a counterexample of the conjecture
that the multipole moments of the spacetime in general relativity is
determined by the multipole integrals. That is, the multipole moments of the
spacetime in general relativity cannot determined by the multipole integrals.

\subsection{Singular source and multipole integral in Newtonian gravity}

We take Newtonian gravity as an example to illustrate the singular source and
the multipole moments.

Solving the source free equation
\begin{equation}
-\nabla^{2}\phi=0,\label{electrostatic}%
\end{equation}
we obtain a special solution
\begin{equation}
\phi_{1}=\frac{Q}{4\pi r}.\label{phi.1}%
\end{equation}
$\phi_{1}$ is singular at $r=0$. The singularity at $r=0$ means a singular
source. For completeness, we provide the procedure of the analysis of the
singularity. Replacing $r$ by $\sqrt{r^{2}+\epsilon^{2}}$, we have
\[
\phi_{1}\left(  \epsilon\right)  =\frac{Q}{4\pi}\frac{1}{\sqrt{r^{2}%
+\epsilon^{2}}}.
\]
Substituting $\phi_{1}\left(  \epsilon\right)  $ into eq. (\ref{electrostatic}%
), we have
\begin{equation}
-\nabla^{2}\phi_{1}\left(  \epsilon\right)  =\frac{3Q}{4\pi}\frac{\epsilon
^{2}}{\left(  r^{2}+\epsilon^{2}\right)  ^{\frac{5}{2}}}.\label{phi.1.limit}%
\end{equation}
Taking the limit $\epsilon\rightarrow0$ on both sides of eq.
(\ref{phi.1.limit}), we have
\[
-\nabla^{2}\phi_{1}=Q\delta\left(  \mathbf{r}\right)  .
\]
There is a Dirac delta source at $r=0$ \cite{jackson1999classical}. The
corresponding monopole moment is
\begin{equation}
Q^{\left(  00\right)  }=\int_{V}d^{3}\mathbf{x}\rho=\int_{-\infty}^{\infty
}\int_{-\infty}^{\infty}\int_{-\infty}^{\infty}dxdydzQ\delta\left(
\mathbf{r}\right)  =Q,
\end{equation}
with $r=\sqrt{x^{2}+y^{2}+z^{2}}$. We can also obtain other solutions from eq.
(\ref{electrostatic}), such as,
\begin{equation}
\phi_{2}=\frac{Q}{4\pi r}+\frac{P}{r^{2}}\cos\theta.\label{phi.2}%
\end{equation}
With the same procedure, we find that the second term in $\phi_{2}$
contributes a finite dipole moment $P$. That is, a solution may possess
different multipole moments simultaneously. The above examples show again that
the multipole moments of the spacetime and the multipole integrals of the
source coincide in Newtonian gravity.

\subsection{Static axial spacetime and Einstein tensor}

The static axial metric can be generally expressed as
\cite{griffiths2009exact,stephani2009exact}
\begin{equation}
ds^{2}=-e^{2\psi}dt^{2}+e^{2\gamma-2\psi}\left(  d\rho^{2}+dz^{2}\right)
+e^{-2\psi}\rho^{2}d\varphi^{2},\label{metric}%
\end{equation}
where $\gamma=\gamma\left(  \rho,z\right)  $ and $\psi=\psi\left(
\rho,z\right)  $. When we calculate the multipole integrals of the source,
spherical coordinates $\left(  r,\theta,\varphi\right)  $ are also used which
is defined as
\begin{align}
\rho &  =r\sin\theta,\nonumber\\
z &  =r\cos\theta.
\end{align}
The Einstein tensor $G_{\mu\nu}=R_{\mu\nu}-\frac{1}{2}\eta_{\mu\nu}R$ with
$\eta_{\mu\nu}=\operatorname*{diag}\left(  -1,1,1,1\right)  $ in the
orthonormal frame are
\begin{align}
G_{tt} &  =2e^{2\psi-2\gamma}\left(  \frac{\partial^{2}\psi}{\partial\rho^{2}%
}+\frac{1}{\rho}\frac{\partial\psi}{\partial\rho}+\frac{\partial^{2}\psi
}{\partial z^{2}}\right)  -G_{\varphi\varphi},\label{G00}\\
G_{\rho\rho} &  =-G_{zz}=e^{2\psi-2\gamma}\left[  \frac{1}{\rho}\frac
{\partial\gamma}{\partial\rho}-\left(  \frac{\partial\psi}{\partial\rho
}\right)  ^{2}+\left(  \frac{\partial\psi}{\partial z}\right)  ^{2}\right]
,\label{G11}\\
G_{\rho z} &  =G_{z\rho}=e^{2\psi-2\gamma}\left(  \frac{1}{\rho}\frac
{\partial\gamma}{\partial z}-2\frac{\partial\psi}{\partial\rho}\frac
{\partial\psi}{\partial z}\right)  ,\label{G12}\\
G_{\varphi\varphi} &  =e^{2\psi-2\gamma}\left[  \left(  \frac{\partial\psi
}{\partial\rho}\right)  ^{2}+\left(  \frac{\partial\psi}{\partial z}\right)
^{2}+\frac{\partial^{2}\gamma}{\partial\rho^{2}}+\frac{\partial^{2}\gamma
}{\partial z^{2}}\right]  .\label{G33}%
\end{align}
For vacuum spacetime, the equation of $\psi$ reads \cite{griffiths2009exact}
\begin{equation}
\frac{\partial^{2}\psi}{\partial\rho^{2}}+\frac{1}{\rho}\frac{\partial\psi
}{\partial\rho}+\frac{\partial^{2}\psi}{\partial z^{2}}=0.\label{equation.ps}%
\end{equation}

\subsection{Singular source and multipole integral in Levi-Civita spacetime}

In this section, we calculate the singular source and multipole integrals in
the Levi-Civita metric.

The Levi-Civita metric \cite{griffiths2009exact} is a static cylindrical
spacetime which reads
\begin{equation}
ds^{2}=-\rho^{2\alpha}dt^{2}+\rho^{2\left(  \alpha^{2}-\alpha\right)  }\left(
d\rho^{2}+dz^{2}\right)  +\rho^{2-2\alpha}d\varphi^{2} \label{metric.c2.x}%
\end{equation}
with
\begin{align}
\psi &  =\alpha\ln\rho,\label{ps.c2}\\
\gamma &  =\alpha^{2}\ln\rho. \label{gamma.c2}%
\end{align}
When $\alpha=-1$, the metric (\ref{metric.c2.x}) becomes
\begin{equation}
ds^{2}=-\frac{1}{\rho^{2}}dt^{2}+\rho^{4}\left(  d\rho^{2}+dz^{2}\right)
+\rho^{4}d\varphi^{2}. \label{metric.c2=-1}%
\end{equation}
Redefining the coordinates $\rho^{2}\rightarrow\left(  \frac{M}{2}\right)
^{\frac{1}{3}}r$, $t\rightarrow2^{\frac{2}{3}}M^{\frac{1}{3}}t$,
$z\rightarrow\left(  \frac{M}{2}\right)  ^{\frac{1}{3}}z$, and $\varphi
\rightarrow\left(  \frac{M}{2}\right)  ^{\frac{1}{3}}\varphi$, the metric
(\ref{metric.c2=-1}) becomes the Kasner metric \cite{geroch1969limits} which
reads
\begin{equation}
ds^{2}=-\frac{2M}{r}dt^{2}+\frac{r}{2M}dr^{2}+r^{2}\left(  dz^{2}+d\varphi
^{2}\right)  . \label{Kasner.metric}%
\end{equation}
When $\alpha=0$ or $1$, the metric (\ref{metric.c2.x}) becomes the Minkowski metric.

Now we calculate the singular source in the Levi-Civita metric
(\ref{metric.c2.x}). Replacing $\rho$ by $\sqrt{\rho^{2}+\epsilon^{2}}$ in
$\psi$ (\ref{ps.c2}) and $\gamma$ (\ref{gamma.c2}) and substituting into the
Einstein tensor $G_{\mu\nu}$, we obtain
\begin{align}
G_{tt}\left(  \epsilon\right)   &  =-\frac{\alpha\left(  \alpha-4\right)
}{\sqrt{g}}\frac{\rho\epsilon^{2}}{\left(  \rho^{2}+\epsilon^{2}\right)  ^{2}%
},\nonumber\\
G_{\rho\rho}\left(  \epsilon\right)   &  =-G_{zz}\left(  \epsilon\right)
=\frac{\alpha^{2}}{\sqrt{g}}\frac{\rho\epsilon^{2}}{\left(  \rho^{2}%
+\epsilon^{2}\right)  ^{2}},\nonumber\\
G_{\varphi\varphi}\left(  \epsilon\right)   &  =\frac{\alpha^{2}}{\sqrt{g}%
}\frac{\rho\epsilon^{2}}{\left(  \rho^{2}+\epsilon^{2}\right)  ^{2}%
}\label{Guv.Kasner}%
\end{align}
with $\frac{1}{\sqrt{g}}=\left(  \rho^{2}+\epsilon^{2}\right)  ^{-\alpha
^{2}+\alpha}\rho$. The energy-momentum tensor is given by
\begin{equation}
T_{\mu\nu}=\frac{1}{8\pi}G_{\mu\nu}=\frac{1}{8\pi}\lim_{\epsilon\rightarrow
0}G_{\mu\nu}\left(  \epsilon\right)  .\label{Guv.definition}%
\end{equation}
The energy-momentum tensor is
\begin{align}
T_{tt} &  =\frac{1}{8}\frac{\alpha\left(  \alpha-4\right)  }{\sqrt{g}}%
\delta\left(  \rho\right)  \delta\left(  \varphi\right)  ,\nonumber\\
T_{\rho\rho} &  =-T_{zz}=\frac{1}{8}\frac{\alpha^{2}}{\sqrt{g}}\delta\left(
\rho\right)  \delta\left(  \varphi\right)  ,\nonumber\\
T_{\varphi\varphi} &  =\frac{1}{8}\frac{\alpha^{2}}{\sqrt{g}}\delta\left(
\rho\right)  \delta\left(  \varphi\right)  .\label{E-M.T.LC}%
\end{align}
In above calculations, $\lim_{\epsilon\rightarrow0}\frac{\epsilon^{2}}{\left(
x^{2}+y^{2}+\epsilon^{2}\right)  }=\pi\delta\left(  x\right)  \delta\left(
y\right)  =\frac{\pi}{\rho}\delta\left(  \rho\right)  \delta\left(
\varphi\right)  $ \cite{ibragimov2009practical} is used. With eq.
(\ref{Multipole.Tensor}), the multipole integrals of the source can be
obtained directly. The monopole integrals of the source in the Levi-Civita
spacetime is
\begin{align}
Q_{tt}^{\left(  00\right)  } &  =-\frac{L_{z}}{8}\alpha^{2}+\frac{L_{z}}%
{2}\alpha,\nonumber\\
Q_{\rho\rho}^{\left(  00\right)  } &  =-Q_{zz}^{\left(  00\right)  }%
=\frac{L_{z}}{8}\alpha^{2},\nonumber\\
Q_{\varphi\varphi}^{\left(  00\right)  } &  =\frac{L_{z}}{8}\alpha
^{2},\nonumber\\
Q_{M}^{\left(  00\right)  } &  =\frac{L_{z}}{2}\alpha,\label{Monopole.LC}%
\end{align}
where $L_{z}\equiv\int_{z_{1}}^{z_{2}}dz$. Other multipole integrals of the
Levi-Civita spacetime vanish. That is, there exists a line source in the
Levi-Civita metric. In ref. \cite{israel1977line}, Israel confirms that
$\alpha$ may be interpreted as the effective gravitational mass per unit
proper length along the $z$ direction \cite{griffiths2009exact}. We provide
the exact result in this paper.

The above results are valid only for $\alpha\in\left(  0,1\right)  $ since
$\varphi\,$can be regarded as a angular coordinate only for $\alpha\in\left(
0,1\right)  $. More details can be found in ref. \cite{griffiths2009exact}.

\subsection{Multipole integral in Weyl spacetime}

In this section, we calculate the multipole integrals in the Weyl spacetime.

The Weyl solution is the general asymptotically flat\ vacuum solution of the
metric (\ref{metric}). In coordinates $\left(  r,\theta,\varphi\right)  $, the
Weyl solution is \cite{griffiths2009exact,stephani2009exact}
\begin{align}
\psi &  =-\sum_{n=0}^{\infty}\frac{a_{n}}{r^{n+1}}P_{n}\left(  \cos
\theta\right)  ,\label{ps.Weyl}\\
\gamma &  =-\sum_{l=0}^{\infty}\sum_{k=0}^{\infty}\frac{\left(  l+1\right)
\left(  k+1\right)  }{\left(  l+k+2\right)  }\frac{a_{l}a_{k}}{r^{l+k+2}%
}\left[  P_{l}\left(  \cos\theta\right)  P_{k}\left(  \cos\theta\right)
-P_{l+1}\left(  \cos\theta\right)  P_{k+1}\left(  \cos\theta\right)  \right]
, \label{gamma.Weyl}%
\end{align}
where $a_{n}$ are expansion coefficients and $P_{n}$ are Legendre polynomials.

The singular source is difficult to calculate. We only calculate the multipole
integrals in the Weyl solution. With the Einstein equation $T_{\mu\nu}%
=\frac{1}{8\pi}G_{\mu\nu}$ and eqs. (\ref{Einstein.static}), we have
\begin{align}
\rho_{M} &  =\frac{1}{8\pi}\left(  G_{tt}+G_{rr}+G_{\theta\theta}%
+G_{\varphi\varphi}\right)  \nonumber\\
&  =\frac{e^{2\psi-2\gamma}}{4\pi r^{2}}\left[  r^{2}\frac{\partial^{2}\psi
}{\partial r^{2}}+2r\frac{\partial\psi}{\partial r}+\frac{\partial^{2}\psi
}{\partial\theta^{2}}+\cot\theta\frac{\partial\psi}{\partial\theta}\right]
.\label{rho.H}%
\end{align}
Replacing $r$ by $\sqrt{r^{2}+\epsilon^{2}}$ in eqs. (\ref{ps.Weyl}) and
(\ref{gamma.Weyl}) and substituting into eq. (\ref{rho.H}), we have
\begin{equation}
\rho_{M}\left(  \epsilon\right)  =\frac{e^{2\psi-2\gamma}}{4\pi r^{2}}%
\sum_{n=0}^{\infty}a_{n}P_{n}\left(  \cos\theta\right)  \left(  r^{2}%
+\epsilon^{2}\right)  ^{-\frac{n+5}{2}}\left[  \left(  2n+3\right)  \left(
n+1\right)  r^{2}\epsilon^{2}+n\left(  n+1\right)  \epsilon^{4}\right]
.\label{rho.H.eps}%
\end{equation}
The multipole integrals of the source reads
\begin{equation}
Q_{M}^{\left(  l0\right)  }=\lim_{\epsilon\rightarrow0}\int dV\rho_{H}\left(
\epsilon\right)  r^{l}Y_{lm}^{\ast}\left(  \theta,\varphi\right)
=a_{l}.\label{Multipole.Weyl}%
\end{equation}
That is, the expansion coefficients $a_{n}$ in the Weyl solution are the
multipole integrals of the source defined in eq. (\ref{Multipole.Scalar}).

\subsection{Singular source and multipole integral in Curzon spacetime}

In this section, we calculate the singular source and multipole integrals in
the Curzon spacetime.

Taking $a_{n}=0$ ($n\geqslant2$) in the Weyl solution, we get the Curzon
solution%
\begin{align}
\psi &  =-\frac{M}{r},\label{ps.Curzon.r}\\
\gamma &  =-\frac{M^{2}}{2}\frac{\sin^{2}\theta}{r^{2}}.
\label{gamma.Curzon.r}%
\end{align}
The Curzon metric in coordinates $\left(  t,r,\theta,\varphi\right)  $ reads
\begin{equation}
ds^{2}=-e^{-\frac{2M}{r}}dt^{2}+e^{-\frac{M^{2}\sin^{2}\theta}{r^{2}}%
+\frac{2M}{r}}\left(  dr^{2}+r^{2}d\theta^{2}\right)  +e^{\frac{2M}{r}}%
r^{2}\sin^{2}\theta d\varphi^{2}. \label{metric.curzon.2}%
\end{equation}
The Curzon metric has same asymptotics with the Schwarzschild metric at
$r\rightarrow\infty$. That is, when $r\rightarrow\infty$, the metric
(\ref{metric.curzon.2}) becomes
\begin{equation}
ds^{2}\sim-\left(  1-\frac{2M}{r}\right)  dt^{2}+\frac{1}{1-\frac{2M}{r}%
}\left(  dr^{2}+r^{2}d\theta^{2}\right)  +r^{2}\sin^{2}\theta d\varphi^{2}.
\end{equation}

We remind that
\begin{equation}
G\left(  r\right)  =-\frac{M}{r}\label{ps.Curzon}%
\end{equation}
is the Green function of the equation (\ref{equation.ps})
\begin{equation}
\frac{\partial^{2}\psi}{\partial z^{2}}+\frac{\partial^{2}\psi}{\partial
\rho^{2}}+\frac{1}{\rho}\frac{\partial\psi}{\partial\rho}=0.
\end{equation}
That is
\begin{equation}
\left(  \frac{\partial^{2}}{\partial z^{2}}+\frac{\partial^{2}}{\partial
\rho^{2}}+\frac{1}{\rho}\frac{\partial}{\partial\rho}\right)  G\left(
r\right)  =4\pi M\delta\left(  x\right)  \delta\left(  y\right)  \delta\left(
z\right)  \label{source.equation}%
\end{equation}
with $r^{2}=\rho^{2}+z^{2}=x^{2}+y^{2}+z^{2}$. This fact indicates that the
Curzon metric is not a vacuum solution at $\rho=0$. Nevertheless, the source
in eq. (\ref{source.equation}) is not the energy-moment in the Einstein equation.

Now we calculate the source in the Curzon solution. Replacing $\rho$ by
$\sqrt{\rho^{2}+\epsilon^{2}}$ in $\psi$ (\ref{ps.Curzon.r}) and $\gamma$
(\ref{gamma.Curzon.r}) and substituting into the Einstein tensor $G_{\mu\nu}$,
we obtain
\begin{align}
G_{tt}\left(  \epsilon\right)   &  =\frac{1}{\sqrt{g}}M\epsilon^{2}%
6\frac{r^{2}\sin\theta}{\left(  r^{2}+\epsilon^{2}\right)  ^{\frac{5}{2}}%
}-G_{\varphi\varphi}\left(  \epsilon\right)  ,\nonumber\\
G_{rr}\left(  \epsilon\right)   &  =-G_{\theta\theta}\left(  \epsilon\right)
=\frac{1}{\sqrt{g}}M^{2}\epsilon^{2}\left[  \frac{\cos^{2}\theta\sin\theta
}{\left(  r^{2}+\epsilon^{2}\right)  ^{2}}+\frac{r^{2}\sin\theta}{\left(
r^{2}+\epsilon^{2}\right)  ^{3}}\right]  ,\nonumber\\
G_{r\theta}\left(  \epsilon\right)   &  =-\frac{1}{\sqrt{g}}M^{2}\epsilon
^{2}\frac{\cos\theta\sin^{2}\theta}{\left(  r^{2}+\epsilon^{2}\right)  ^{2}%
},\nonumber\\
G_{\varphi\varphi}\left(  \epsilon\right)   &  =\frac{1}{\sqrt{g}}%
M^{2}\epsilon^{2}\left[  -\frac{\sin\theta\cos\left(  2\theta\right)
}{\left(  r^{2}+\epsilon^{2}\right)  ^{2}}-\frac{2r^{2}\sin\theta\cos\left(
2\theta\right)  }{\left(  r^{2}+\epsilon^{2}\right)  ^{3}}+\frac{r^{2}%
\sin\theta}{\left(  r^{2}+\epsilon^{2}\right)  ^{3}}\right]
.\label{EinsteinTensor.Curzon}%
\end{align}
Taking the limit $\epsilon\rightarrow0$, we obtain the energy-momentum tensor%
\begin{align}
T_{tt} &  =\frac{M}{\sqrt{g}}\delta\left(  r\right)  \delta\left(
\theta\right)  \delta\left(  \varphi\right)  -T_{\varphi\varphi},\nonumber\\
T_{rr} &  =-T_{\theta\theta}=\frac{1}{\sqrt{g}}\frac{M^{2}}{6r}\left(
1+3\cos^{2}\theta\right)  \delta\left(  r\right)  \delta\left(  \theta\right)
\delta\left(  \varphi\right)  ,\nonumber\\
T_{r\theta} &  =-\frac{1}{\sqrt{g}}\frac{M^{2}}{2r}\sin\theta\cos\theta
\delta\left(  r\right)  \delta\left(  \theta\right)  \delta\left(
\varphi\right)  ,\nonumber\\
T_{\varphi\varphi} &  =-\frac{1}{\sqrt{g}}\frac{M^{2}}{3r}\left(  -3+5\cos
^{2}\theta\right)  \delta\left(  r\right)  \delta\left(  \theta\right)
\delta\left(  \varphi\right)  .\label{EnergyMomentum.Curzon}%
\end{align}
with $r^{2}=x^{2}+y^{2}+z^{2}$. In above calculations, following results are
used \cite{chen2017singular}%
\begin{align*}
\lim_{\epsilon\rightarrow0}\frac{\epsilon^{2}}{\left(  r^{2}+\epsilon
^{2}\right)  ^{\frac{5}{2}}} &  =\frac{4\pi}{3}\delta\left(  x\right)
\delta\left(  y\right)  \delta\left(  z\right)  ,\\
\lim_{\epsilon\rightarrow0}\frac{\epsilon^{2}}{r^{2}\left(  r^{2}+\epsilon
^{2}\right)  ^{\frac{3}{2}}} &  =4\pi\delta\left(  x\right)  \delta\left(
y\right)  \delta\left(  z\right)  ,\\
\lim_{\epsilon\rightarrow0}\frac{r}{\left(  r^{2}+\epsilon^{2}\right)
^{\frac{1}{2}}} &  =1,\\
\delta\left(  x\right)  \delta\left(  y\right)  \delta\left(  z\right)   &
=\frac{1}{r^{2}\sin\theta}\delta\left(  r\right)  \delta\left(  \theta\right)
\delta\left(  \varphi\right)  .
\end{align*}
With eq. (\ref{Multipole.Tensor}), the multipole integrals of the source can
be obtained directly. The multipole integrals in the Curzon spacetime read
\begin{align}
Q_{tt}^{\left(  00\right)  } &  =M,\nonumber\\
Q_{\mu\nu}^{\left(  10\right)  } &  =0,\nonumber\\
Q_{M}^{\left(  00\right)  } &  =M.
\end{align}
Other multipole integrals vanish.  

The first term in $T_{00}$ is a same Dirac delta source with the Schwarzschild
metric \cite{chen2017singular}, which is the reason that the Curzon metric has
the same asymptotics with the Schwarzschild metric at infinity. The
energy-momentum tensor in the Curzon spacetime possess other polarized terms
diverging stronger that the Dirac delta source, which is the reason that the
Curzon metric is not spherically symmetric and the singularity in Curzon
metric have different asymptotics along different directions. The asymptotics
of the Curzon spacetime are analyzed in ref \cite{scott1986curzon}. In ref.
\cite{bonnor1992physical}, the author mentioned that the far-field of the
Curzon metric is of a mass at $r=0$ with the multipole moments on it. We
provide the quantitative\ results.

\subsection{Singular source and source integral in Schwarzschild spacetime}

In a previous work \cite{chen2017singular}, we provide the singular source in
the Schwarzschild spacetime. The metric of the Schwarzschild spacetime reads
\[
ds^{2}=-\left(  1-\frac{2M}{r}\right)  dt^{2}+\left(  1-\frac{2M}{r}\right)
^{-1}dr^{2}+r^{2}\left(  d\theta^{2}+d\phi^{2}\right)  .
\]

For convenience of discussion, we present the singular source and the
multipole integrals of the Schwarzschild spacetime here. The singular source
in the Schwarzschild spacetime reads \cite{chen2017singular}
\begin{align}
T_{tt} &  =\frac{M}{r^{2}\sin\theta}\delta\left(  r\right)  \delta\left(
\theta\right)  \delta\left(  \phi\right)  ,\nonumber\\
T_{rr} &  =-\frac{M}{r^{2}\sin\theta}\delta\left(  r\right)  \delta\left(
\theta\right)  \delta\left(  \phi\right)  ,\nonumber\\
T_{\theta\theta} &  =T_{\phi\phi}=\frac{M}{2r^{2}\sin\theta}\delta\left(
r\right)  \delta\left(  \theta\right)  \delta\left(  \phi\right)
.\label{EnergyMomentum.Schwarzschild}%
\end{align}
The monopole integrals of the source in the Schwarzschild spacetime are
\begin{align}
Q_{tt}^{\left(  00\right)  } &  =M,\nonumber\\
Q_{rr}^{\left(  00\right)  } &  =-M,\nonumber\\
Q_{\theta\theta}^{\left(  00\right)  } &  =Q_{\varphi\varphi}^{\left(
00\right)  }=\frac{M}{2},\nonumber\\
Q_{M}^{\left(  00\right)  } &  =M.\label{Monopole.Schwarzschild}%
\end{align}

All above results about the multipole moments of the source are used to
discuss the multipole moments problem in general relativity.

\subsection{Multipole moment problem in general relativity}

In this section, we illustrate that the multipole moments of the gravitational
field in general relativity cannot be determined by the multipole integrals of
the source.

In above sections, we show that the Levi-Civita spacetime, the Curzon
spacetime and the Schwarzschild spacetime only have nonvanishing monopole
moments of the source. However, the energy-momentum tensor $T_{\mu\nu}$ in
these spacetimes are quite different. The Levi-Civita spacetime possesses a
line-like source, while the Levi-Civita spacetime and the Schwarzschild
spacetime possess point-like sources. Furthermore, the Levi-Civita spacetime
and the Schwarzschild spacetime possess the same multipole integrals of the
sources. Nevertheless, they have different expansions at infinity. The above
results show that the multipole moments of the spacetime cannot be determined
by multipole integrals of the source.

Generally, the radial coordinate $r$ in curved spacetimes is difficult to
define so that the multipole integrals of the source is difficult to define.
Nevertheless, the curzon spacetime and the Schwarzschild spacetime only
possess nonvanishing $Q_{M}^{\left(  00\right)  }$ which is independent of the
definition of $r$. That is, the conclusion in this paper is independent of the
definition of the radial coordinate $r$ and the results is invariant under the
space coordinates transform.

\section{Conclusion and outlook}

In Newtonian gravity, the multipole moments of the gravitational field is
determined by the source distribution --- a series of multipole integrals. In
this paper, with a counterexample, we show that the multipole moments of the
gravitational field cannot be determined by multipole integrals of the source
in general relativity. The Curzon spacetime and the Schwarzschild spacetime
possess same multipole integrals of the source, while they possess different
multipole moments of the spacetime.

The multipole moments of the spacetime are widely discussed, while the
multipole integrals are not. In this paper, we show that there exist singular
sources at singularities in static axial spacetimes. Outside the
singularities, the spacetimes are vacuum solutions of the Einstein equation.
Nevertheless, point-like or line-like sources exist at singularities. These
singular sources contribute to multipole integrals which allow us to discuss
the multipole moments problem in general relativity. Besides, we provide a
useful method to analyze the singular source quantitatively. The validity of
the method is verified by the singular source and multipole integrals in the
Levi-Civita spacetime and our previous work \cite{chen2017singular}.

In the follow contents, we discuss some open questions concerning the singular
source and the multipole moments in general relativity.

A singular source indicates whether the corresponding singularity is
physically acceptable or not. We define a delta function in curved spacetime
\[
\delta^{3}\left(  x\right)  =\frac{1}{\sqrt{g}}\delta\left(  x_{1}%
-x_{1}^{\prime}\right)  \delta\left(  x_{2}-x_{2}^{\prime}\right)
\delta\left(  x_{3}-x_{3}^{\prime}\right)  ,
\]
where $x_{i}$ and $x_{i}^{\prime}$ ($i=1,2,3$) are space coordinates. The
physically acceptable mass at one point should be finite in total. That is,
only singularities diverging slower than $\delta^{3}\left(  x\right)  $ are
physically acceptable, while singularities diverging stronger than $\delta
^{3}\left(  x\right)  $ are not physically acceptable. In this paper, the
singularity in the Levi-Civita spacetime and the Schwarzschild spacetime are
physically acceptable, while the singularity in the Curzon spacetime are not
physically acceptable.

There are other viewpoint about the multipole moments of the spacetime. We
illustrate with the electrostatic field since the electrostatic field
satisfies the same equation with Newtonian gravity. In the electrostatic
field, the multipole moments can be defined as the multipole integrals. The
fact that the multipole moments coincides with the multipole integrals can be
explained as that the electrostatic field itself does contribute to the
multipole moments. The multipole integrals is about the integrals over the
electric charge. The electrostatic field does not carry any electric charge so
that the electrostatic field does not contribute to the multipole moments.
However, general relativity is different with the electrostatic field. The
charge of general relativity is the mass. The gravitational field must
contribute to the multipole moments of the spacetime if we suppose that the
gravitational field possesses the energy. In this viewpoint, the definition of
the energy of the gravitational field is a subproblem of the definition of the
multipole moments of the spacetime. In Thorne's work
\cite{thorne1980multipole}, the multipole moments of the spacetime involves
the definition of the energy-moment of the gravitational field --- the Landau
pseudo-tensor. With this viewpoint, we provide another method to calculate the
multipole integrals of the source in general relativity.

We demonstrate that the gravitational field itself contributes to the
multipole moments of the spacetime in above paragraph. Another problem is
whether the gravitational field itself without any source or any singularity
(the gravitational wave) collapses into a asymptotically flat spacetime or not.

\section*{Acknowledgments}

We are very indebted to Dr G. Zeitrauman for his encouragement. This work is
supported in part by Nankai Zhide foundation and NSF of China under Grant
No.11575125 and No.11675119.

\section{Appendix}

In the appendix, we provide another method to calculate the multipole
integrals of the source in the spacetime.

The multipole moments of the spacetime are also expressed as volume integrals
\cite{hernandez2016source}
\begin{align}
M^{\left(  lm\right)  } &  =\int dV\rho_{M}r^{l}Y_{lm}^{\ast}\left(
\theta,\varphi\right)  -\frac{1}{4\pi}\int dVD_{i}D^{i}\left(  r^{l}%
Y_{lm}^{\ast}\left(  \theta,\varphi\right)  \right)  \nonumber\\
&  =Q_{M}^{\left(  lm\right)  }+Q_{F}^{\left(  lm\right)  }%
\label{Multipole.Hernandez}%
\end{align}
where $Q_{F}^{\left(  lm\right)  }=-\frac{1}{4\pi}\int dVD_{i}D^{i}\left(
r^{l}Y_{lm}^{\ast}\left(  \theta,\varphi\right)  \right)  \ $with the
subscript "$F$" of $Q_{F}^{\left(  lm\right)  }$ means "field", and
$dV\equiv\sqrt{g}dx^{1}dx^{2}dx^{3}$ with $g=\xi^{2}\det g_{ij}$
(\ref{Einstein.static}). The multipole moments of the spacetime in eq.
(\ref{Multipole.Hernandez}) coincide, at least in the axial case, with the
multipole moments of the spacetime defined by Thorne or Geroch
\cite{hernandez2016source}. The Multipole moments of the spacetime in eq.
(\ref{Multipole.Hernandez}) are exact differentials so that they can be
converted to boundary integrals \cite{hernandez2016source}
\begin{equation}
M^{\left(  lm\right)  }=\frac{1}{4\pi}\int_{\partial V}dS\left[  D^{i}\xi
r^{l}Y_{lm}^{\ast}\left(  \theta,\varphi\right)  -\xi D^{i}\left(  r^{l}%
Y_{lm}^{\ast}\left(  \theta,\varphi\right)  \right)  \right]
.\label{Multipole.boundary}%
\end{equation}

There are two parts in $M^{\left(  lm\right)  }$: multipole integrals
$Q_{M}^{\left(  lm\right)  }$ and source free integrals $Q_{F}^{\left(
lm\right)  }$. With the viewpoint in the conclusion, $Q_{M}^{\left(
lm\right)  }$ and $Q_{F}^{\left(  lm\right)  }$ can be regarded as the
multipole moments contribute by the source and the gravitational field
respectively. When we take the first order of the asymptotically flat
approximation, $D_{i}$ is replaced by $\partial_{i}$. In this case,
$Q_{F}^{\left(  lm\right)  }$ vanishes, which means that $M^{\left(
lm\right)  }$ is totally contributed by $Q_{M}^{\left(  lm\right)  }$. That
is, we replace $D_{i}$ with $\partial_{i}$, eq. (\ref{Multipole.boundary}) and
obtain the source integrals $Q_{M}^{\left(  lm\right)  }$. In ref.
\cite{hernandez2016source}, the author shows that in the Weyl solution,
replacing $D_{i}$ with $\partial_{i}$ in eq. (\ref{Multipole.boundary})
provides the same multipole integrals with the results we calculated in
section 3.

%(正文结束)――――――――――――――――――――――――――――――――――――――――――――――――――

%%%%%%%%%%%%%%%%%%%参考文献%%%%%%%%%%%%%%%%%%%%%%%%%%%

%%%%%%%%%%%%%%%%%%%bibtex形式的参考文献%%%%%%%%%%%%%%%
%\bibliographystyle{JHEP} %参考文献的风格(.bst)
%\bibliography{refs} %参考文献文件(.bib)
%\nocite{*} %若不去掉注释，没有被引用的文献也被列出

%%%%%%%%%%%%%%%%%%%bbl形式的参考文献%%%%%%%%%%%%%%%%%%

%%%%%%%%%%%%%%%%%%%%%%%%%%%%%%%%%%%%%%%%%%%%%%%%%%%%%%

\end{CJK*}
\end{document}